\documentclass[11pt,twoside]{article}

\usepackage{asp2014}

\aspSuppressVolSlug
\resetcounters

\begin{document}

\title{Building Trustworthy Machine Learning Models for Astronomy}

\author{Michelle Ntampaka$^{1, 2}$, Matthew Ho$^{3,4}$, and Brian Nord$^{5, 6, 7}$}
\affil{$^1$Data Science Mission Office, Space Telescope Science Institute, Baltimore, MD 21218, USA; \email{mntampaka@stsci.edu}}
\affil{$^2$Johns Hopkins University, Baltimore, MD 21218, USA}
\affil{$^3$McWilliams Center for Cosmology, Department of Physics, Carnegie Mellon University, Pittsburgh, PA 15213, USA}
\affil{$^4$NSF AI Planning Institute for Physics of the Future, Carnegie Mellon University, Pittsburgh, PA 15213, USA}
\affil{$^5$Fermi National Accelerator Laboratory, Batavia, IL 60510, USA}
\affil{$^6$Department of Astronomy and Astrophysics, University of Chicago, IL 60637, USA}
\affil{$^7$Kavli Institute for Cosmological Physics, University of Chicago, Chicago, IL 60637, USA}

\paperauthor{Michelle Ntampaka}{mntampaka@stsci.edu}{0000-0002-0144-387X}{Space Telescope Science Institute}{Data Science Mission Office}{Baltimore}{MD}{21218}{USA}

%\aindex{Ntampaka,~M.}
%\aindex{Ho,~M.}
%\aindex{Author3,~S.}

\begin{abstract}
Astronomy is entering an era of data-driven discovery, due in part to modern machine learning (ML) techniques enabling powerful new ways to interpret observations. 
This shift in our scientific approach requires us to consider whether we can trust the black box.  
Here, we overview methods for an often-overlooked step in the development of ML models: building community trust in the algorithms.  
Trust is an essential ingredient not just for creating more robust data analysis techniques, but also for building confidence within the astronomy community to embrace machine learning methods and  results.
\end{abstract}

\section{Introduction}

Astronomy is becoming a data-driven field.  
Upcoming surveys will produce exquisitely detailed observations, and we will need to develop appropriate techniques to draw meaningful conclusions from these rich data sets.  The next generation of astronomers will need to be bilingual \textemdash{} fluent in the scientific questions of astronomy and also in modern data analysis techniques, such as machine learning (ML).

%Let's try it without this paragraph and see where it takes us!
%Astronomy is the ideal sandbox for machine learning.   Astronomers have a culture of making our data publicly available, and we have rich and complex observations to answer well-posed questions. Our observations are typically non-monetizeable and cannot be used to advertise or sell a product.  Astronomical data have minimal privacy concerns, in contrast with some other scientific disciplines, where adhering to data protection and privacy requirements is an ethical challenge  \citep[e.g., research pertaining to healthcare records,][]{vayena2018machine}.   And while these benefits do not exempt astronomers from ethical concerns \citep[such as those described in][]{o2016weapons}, astronomers enjoy tremendous freedom to explore our data and experiment with interpretation in a low-stakes environment.

Astronomical data sets capture complex, non-linear phenomena where conclusions must often be drawn from limited information; this makes astronomy an attractive sandbox for testing new machine learning algorithms.  While astronomy data are well-suited for exploring ML techniques,  it is worth contemplating whether the converse is true; we should carefully consider whether machine learning is the right tool for astronomy.  
As our field grows in ML fluency and experience, the most immediate question we should be asking is: Can ML be trusted?

\section{Three Tools for Building Trust}

Machine learning methods can make remarkable improvements in data interpretation, and they are often treated as impenetrable black boxes, but this need not be the case.  Using ML as a black box is dangerous in scientific endeavors and can lead us to draw incorrect conclusions, but there are tools that we can use to build trust in our models and results.

\subsection{Scrutinize All Results}

It is widely known that ML algorithms and tools, like any other set of computational modeling schemes, are not automatically trustworthy. 
Moreover, significant peril accompanies their use in contexts that require unbiased or quantifiably precise results: these circumstances are well known to occur in both science and society \citep[e.g.,][]{2021arXiv210502637G, 2021arXiv211100961C,  2019arXiv191207376B, 2021arXiv210615590B}. 
It is imperative that ML practitioners proceed with caution.
In astronomical literature, there exist relatively few discussions on uncertainty quantification and bias. 
Nevertheless, to achieve reliable scientific outcomes, all ML applications must be rigorously scrutinized. 

% Reliable and just outcomes for applications in both science and society depend on 
% and challenges in using ML techniques without considering downstream outcomes --- e.g., ..., ... ,... (cite, cite, cite). 
% Experienced ML practitioners know that ML algorithms cannot be trusted blindly, and astronomers who use these techniques must proceed with caution.  
% When a machine learning model produces results that are much better or much worse than expected, those results should be carefully scrutinized.

\citet{2019NatCo..10.1096L} presents several case studies of ML algorithms that produce excellent results, but are caught ``cheating'' because they had learned to exploit spurious correlations or unphysical data artifacts. 
In their Figure 2, they showed a surprising result from interpreting one particular trained image classifier:  the model had learned to identify a horse in natural images, but not by looking for the shape of the nose or the presence of a tail, or even by searching for a background of rolling green hills.  
This particular image classifier had learned to identify the copyright stamp of an equine photographer in order to label an image as containing a horse.  
More worrisome, artificially adding this stamp to another image (in their example, of a red convertible), would cause that image to also be classified as containing a horse. 

Astronomers should learn from this case study:  if we are to build trustworthy ML models to interpret our valuable data sets, it is essential that we identify and mitigate these kinds of cheating behaviors.  
Before our models are used to draw inferences, the results should be carefully scrutinized.

\subsection{Investigate Models}

Attempting to understand how a trained ML model works is an essential step for building trust in the model. 
Methodology for investigating the behavior of ML models is an active development frontier in both academic and industry research. 
Simple, interpretable models such as EBMs \citep{lou2013accurate} are commonly used to enable flexible learning on par with state-of-the art black box models while ensuring full intelligibility through measurable feature importances. 
Diagnostic tools such as LIME \citep{ribeiro2016should} and SHAP \citep{lundberg2017unified} allow users to find decision boundaries within pre-trained black box models and gain insight as to how the models are making predictions. 
Explanation procedures such as DiCE \citep{mothilal2020explaining} provide algorithms for generating counterfactual examples for existing model predictions, i.e. ``what-if'' scenarios which dictate how slightly different inputs might drastically impact outputs. 
Each of these techniques allow users to peek into the inner workings of ML models and diagnose their robustness and reliability in practical applications.

Complex deep models, such as convolutional neural networks, tend to be the most difficult to interpret, but there are interrogation schemes for even these.  
For example, model visualization \citep[e.g.,][]{cnn_see} can give some insight into the relevant input features and is a useful check for whether the model is analyzing data similarly to a human expert, or in a novel way.  
Saliency maps \citep{2013arXiv1312.6034S} are powerful tools for identifying which input pixels are the most informative, and adversarial attacks \citep[e.g.,][]{2017arXiv171209665B} can provide insight on what kinds of inputs are easily misinterpreted.

It is no longer true that deep ML models must be treated as black boxes, and these are just a sample of the types of interpretation schemes that can shed light on how and why the model draws conclusions from data.  
Though they may not result in complete understanding of the model's inner workings, applying one or more investigative techniques is absolutely essential for building community trust in machine learning.

\subsection{Use Concise Language}

Every field has jargon and terminology that allow experts to quickly convey complex ideas.  
Interdisciplinary research comes with the challenge of adopting two distinct sets of vocabulary and merging them into a single, coherent dialect that can be quickly understood by experts in multiple fields.  
To build trust in ML models, astronomers should be should be aware of linguistic differences between ML and astronomy, and should take care to use technical words with precision.

For example, one word that is commonly misapplied in astronomical literature is ``overfit.''  
This word is precisely defined in the field of machine learning:  an overfit model has learned the training data too well.  
It has not just learned the trends, but it has also learned the noise of the training data.  
While an overfit model will give stunning results on the training data, it will underperform on the test data.  
Compared to a model that is properly fit, an overfit model will give larger errors on the test data.

Astronomers should not use ``overfit'' to describe a very different failure mode: when a model has identified simulation-specific details that are not true of reality.  
If a model is heavily relying on nonphysical details (for example, simulation artifacts), the model may not generalize to observational data.  
To describe this scenario, Ntampaka and Vikhlinin (under review) 
%\citet{nvinprep} 
propose the word ``overspecialize.''  
An overspecialized model may produce very good results on the simulation data, but the model will not generalize to observational data.  
The image classifier described in the previous subsection is an example of overspecialization:  the copyright stamp is akin to a simulation artifact, and while it may appear to be a useful piece of information, a model trained to rely on it will give suboptimal results on other data sets.

\section{Conclusion}

Machine learning is often peddled as a way of making order-of-magnitude improvements at the cost of interpretability.  
Black box models, however, are not only scientifically unsatisfying, but they can often lead us to draw dangerous or incorrect conclusions.   
As our field develops ML expertise, we should invest in learning to build trustworthy machine learning models by scrutinizing results, investigating models, and presenting our research with concise language.
Rigorously investigated ML models are tremendously beneficial; they have the potential to be powerful tools for facilitating data-driven astronomical discovery.

\acknowledgements MN and BN would like to thank the organizers and attendees of the Machine Learning Tools for Research in Astronomy conference in Ringberg, Germany, for the many hours of productive conversation on how to build trustworthy machine learning models.

\bibliography{references}  

\begin{thebibliography}{}
\expandafter\ifx\csname natexlab\endcsname\relax\def\natexlab#1{#1}\fi
\expandafter\ifx\csname url\endcsname\relax
  \def\url#1{\texttt{#1}}\fi
\expandafter\ifx\csname urlprefix\endcsname\relax\def\urlprefix{URL }\fi
\providecommand{\eprint}[2][]{\url{#2}}

\bibitem[{{Brown} et~al.(2017){Brown}, {Man{\'e}}, {Roy}, {Abadi}, \&
  {Gilmer}}]{2017arXiv171209665B}
{Brown}, T.~B., {Man{\'e}}, D., {Roy}, A., {Abadi}, M., \& {Gilmer}, J. 2017,
  arXiv e-prints, arXiv:1712.09665. \eprint{1712.09665}

\bibitem[{Chollet(2016)}]{cnn_see}
Chollet, F. 2016, How convolutional neural networks see the world.
  \urlprefix\url{https://blog.keras.io/how-convolutional-neural-networks-see-the-world.html}

\bibitem[{Feldscher(2021)}]{redblue}
Feldscher, J. 2021, Politico

\bibitem[{{Lapuschkin} et~al.(2019){Lapuschkin}, {W{\"a}ldchen}, {Binder},
  {Montavon}, {Samek}, \& {M{\"u}ller}}]{2019NatCo..10.1096L}
{Lapuschkin}, S., {W{\"a}ldchen}, S., {Binder}, A.~e., {Montavon}, G., {Samek},
  W., \& {M{\"u}ller}, K.-R. 2019, Nature Communications, 10, 1096.
  \eprint{1902.10178}

\bibitem[{{Ntampaka} \& {Vikhlinin}(view)}]{nvinprep}
{Ntampaka}, M., \& {Vikhlinin}, A. {under review}

\bibitem[{O'Neil(2016)}]{o2016weapons}
O'Neil, C. 2016, Weapons of math destruction: How big data increases inequality
  and threatens democracy (Crown)

\bibitem[{{Simonyan} et~al.(2013){Simonyan}, {Vedaldi}, \&
  {Zisserman}}]{2013arXiv1312.6034S}
{Simonyan}, K., {Vedaldi}, A., \& {Zisserman}, A. 2013, ArXiv e-prints,
  arXiv:1312.6034. \eprint{1312.6034}

\bibitem[{Vayena et~al.(2018)Vayena, Blasimme, \& Cohen}]{vayena2018machine}
Vayena, E., Blasimme, A., \& Cohen, I.~G. 2018, PLoS medicine, 15, e1002689

\end{thebibliography}


\begin{thebibliography}{}
\expandafter\ifx\csname natexlab\endcsname\relax\def\natexlab#1{#1}\fi
\expandafter\ifx\csname url\endcsname\relax
  \def\url#1{\texttt{#1}}\fi
\expandafter\ifx\csname urlprefix\endcsname\relax\def\urlprefix{URL }\fi
\providecommand{\eprint}[2][]{\url{#2}}

\bibitem[{{Birhane} \& {Cummins}(2019)}]{2019arXiv191207376B}
{Birhane}, A., \& {Cummins}, F. 2019, arXiv e-prints, arXiv:1912.07376.
  \eprint{1912.07376}

\bibitem[{{Birhane} et~al.(2021){Birhane}, {Kalluri}, {Card}, {Agnew}, {Dotan},
  \& {Bao}}]{2021arXiv210615590B}
{Birhane}, A., {Kalluri}, P., {Card}, D., {Agnew}, W., {Dotan}, R., \& {Bao},
  M. 2021, arXiv e-prints, arXiv:2106.15590. \eprint{2106.15590}

\bibitem[{{Brown} et~al.(2017){Brown}, {Man{\'e}}, {Roy}, {Abadi}, \&
  {Gilmer}}]{2017arXiv171209665B}
{Brown}, T.~B., {Man{\'e}}, D., {Roy}, A., {Abadi}, M., \& {Gilmer}, J. 2017,
  arXiv e-prints, arXiv:1712.09665. \eprint{1712.09665}

\bibitem[{Chollet(2016)}]{cnn_see}
Chollet, F. 2016, How convolutional neural networks see the world.
  \urlprefix\url{https://blog.keras.io/how-convolutional-neural-networks-see-the-world.html}

\bibitem[{{{\'C}iprijanovi{\'c}} et~al.(2021){{\'C}iprijanovi{\'c}}, {Kafkes},
  {Perdue}, {Pedro}, {Snyder}, {S{\'a}nchez}, {Madireddy}, {Wild}, \&
  {Nord}}]{2021arXiv211100961C}
{{\'C}iprijanovi{\'c}}, A., {Kafkes}, D., {Perdue}, G.~N., {Pedro}, K.,
  {Snyder}, G., {S{\'a}nchez}, F.~J., {Madireddy}, S., {Wild}, S.~M., \&
  {Nord}, B. 2021, arXiv e-prints, arXiv:2111.00961. \eprint{2111.00961}

\bibitem[{{Griffiths} et~al.(2021){Griffiths}, {Schwaller}, \&
  {Lee}}]{2021arXiv210502637G}
{Griffiths}, R.-R., {Schwaller}, P., \& {Lee}, A.~A. 2021, arXiv e-prints,
  arXiv:2105.02637. \eprint{2105.02637}

\bibitem[{{Lapuschkin} et~al.(2019){Lapuschkin}, {W{\"a}ldchen}, {Binder},
  {Montavon}, {Samek}, \& {M{\"u}ller}}]{2019NatCo..10.1096L}
{Lapuschkin}, S., {W{\"a}ldchen}, S., {Binder}, A.~e., {Montavon}, G., {Samek},
  W., \& {M{\"u}ller}, K.-R. 2019, Nature Communications, 10, 1096.
  \eprint{1902.10178}

\bibitem[{Lou et~al.(2013)Lou, Caruana, Gehrke, \& Hooker}]{lou2013accurate}
Lou, Y., Caruana, R., Gehrke, J., \& Hooker, G. 2013, in Proceedings of the
  19th ACM SIGKDD international conference on Knowledge discovery and data
  mining, 623

\bibitem[{Lundberg \& Lee(2017)}]{lundberg2017unified}
Lundberg, S.~M., \& Lee, S.-I. 2017, in Proceedings of the 31st international
  conference on neural information processing systems, 4768

\bibitem[{Mothilal et~al.(2020)Mothilal, Sharma, \&
  Tan}]{mothilal2020explaining}
Mothilal, R.~K., Sharma, A., \& Tan, C. 2020, in Proceedings of the 2020
  Conference on Fairness, Accountability, and Transparency, 607

\bibitem[{Ribeiro et~al.(2016)Ribeiro, Singh, \& Guestrin}]{ribeiro2016should}
Ribeiro, M.~T., Singh, S., \& Guestrin, C. 2016, in Proceedings of the 22nd ACM
  SIGKDD international conference on knowledge discovery and data mining, 1135

\bibitem[{{Simonyan} et~al.(2013){Simonyan}, {Vedaldi}, \&
  {Zisserman}}]{2013arXiv1312.6034S}
{Simonyan}, K., {Vedaldi}, A., \& {Zisserman}, A. 2013, ArXiv e-prints,
  arXiv:1312.6034. \eprint{1312.6034}

\end{thebibliography}

\end{document}